# Detecting Lyme Disease Using Antibody-Functionalized Single-Walled Carbon Nanotube Transistors

Author names and affiliations:
Mitchell B. Lerner*[1], Jennifer Dailey*[1], Brett R. Goldsmith[1#], Dustin Brisson[2], A.T. Charlie Johnson[1]
* These authors contributed equally to this work
[1] Department of Physics and Astronomy, University of Pennsylvania, 209 South 33rd Street, Philadelphia  PA 19104

[2] Department of Biology, University of Pennsylvania, 3740 Hamilton Walk, Philadelphia PA 19104
[#] Current address SPAWAR Systems Center Pacific, 53560 Hull St., San Diego  CA  92152

Corresponding author: A.T. Charlie Johnson, cjohnson@physics.upenn.edu

Present/permanent address:  Department of Physics and Astronomy, University of Pennsylvania, 209 South 33rd Street, Philadelphia, PA 19104

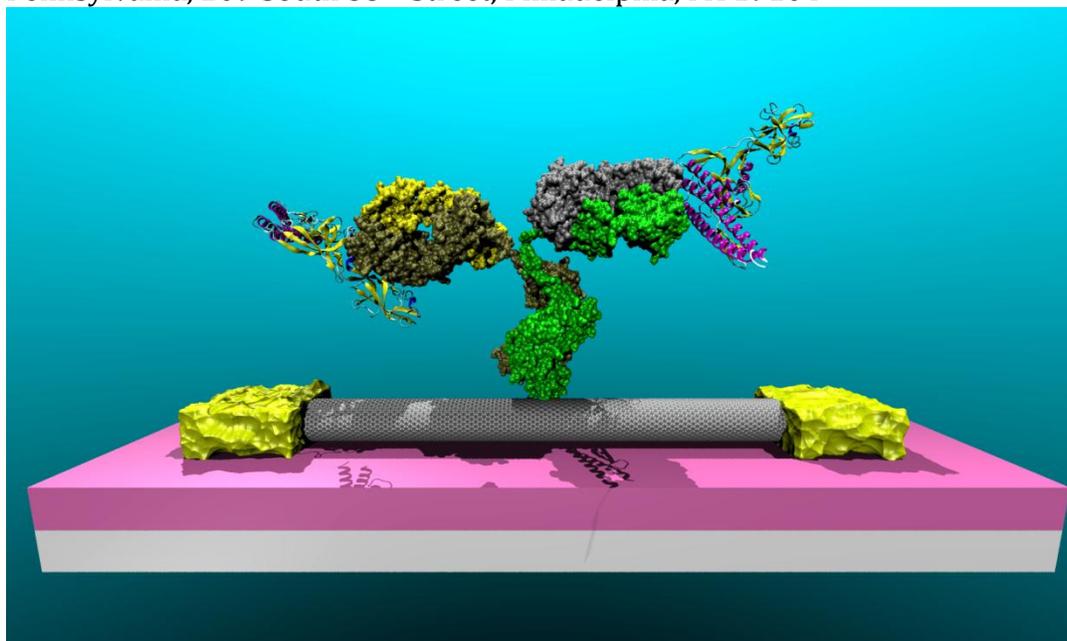

**Abstract:**
We examined the potential of antibody-functionalized single-walled carbon nanotube (SWNT) field-effect transistors (FETs) for use as a fast and accurate sensor for a Lyme disease antigen.  Biosensors were fabricated on oxidized silicon wafers using chemical vapor deposition grown carbon nanotubes that were functionalized using diazonium salts.  Attachment of Borrelia burgdorferi (Lyme) flagellar antibodies to the nanotubes was verified by Atomic Force Microscopy and electronic measurements.  A reproducible shift in the turn-off voltage of the semiconducting SWNT FETs was seen upon incubation with Borrelia burgdorferi flagellar antigen, indicative of the nanotube FET being locally gated by the residues of flagellar protein bound to the antibody.  This sensor effectively detected antigen in buffer at concentrations as low as 1 ng/ml, and the response varied strongly over a concentration range coinciding with levels of clinical interest.  Generalizable binding chemistry gives this biosensing platform the potential to be expanded to monitor other relevant antigens, enabling a multiple vector sensor for Lyme disease.  The speed and sensitivity of this biosensor make it an ideal candidate for development as a medical diagnostic test.

Keywords: carbon nanotubes  Lyme FET

## 1. Introduction

Lyme disease is a tick-borne illness caused by the bacteria Borrelia burgdorferi, generating at least 30,000 new cases in the United States each year, although there are likely many more cases that go undetected or misdiagnosed due to generality of symptoms (Centers for Disease Control and Prevention 2011).  Of the patients diagnosed with Lyme disease, many are originally misdiagnosed due to the general symptoms of the disease (Williams et al. 1990), inconsistent disease presentation in patients (Aguerorosenfeld et al. 1993), and lack of sensitive testing available for early stages of the infection (Bakken et al. 1997).  Late detection of Lyme disease can result in further complications including arthritis and permanent neurological disorders (Marques 2008). <u>Diagnosis of Lyme disease is severely hindered by the lack of reliable diagnostic tools</u> despite its importance to treatment success (Murray and Shapiro 2010; O'Connell 2010). A reliable and rapid laboratory diagnostic tool is crucial to reducing the number of misdiagnosed patients and for investigating appropriate treatment protocols for chronic Lyme disease.

In recent years, great progress has been made in the field of carbon nanotube field effect transistor (CNT FET)-based biosensors (Allen et al. 2007).  Benefits of nanotube-based sensors include the speed and reliability obtained from performing multiple assays in parallel (Chikkaveeraiah et al. 2009).  Protein-functionalized nanotube-based FETs are of great research and clinical interest for several reasons.  Their nanometer size is comparable to the size of many biomolecules of interest, suggesting a unique biocompatible platform (Harrison and Atala 2007; Lerner et al. 2011; Sudibya et al. 2009).  Additionally, since every atom of a carbon nanotube is located on its surface, in direct contact with the environment, they are a clear choice for direct environmental sensing. Commercially available assays for Lyme-specific antigens in urine and cerebrospinal fluid have a limit of detection of 12-15 ng/mL (Shah et al. 2004). We hypothesized that an antibody-functionalized SWNT FET immunosensor would be able to detect the small amount of Lyme antigen that is present in bodily fluids at very early stages of Borrelia infection (Harris and Stephens 1995) since protein-functionalized nano-enabled sensors have demonstrated very low detection limits, on the order of fM (Duan et al. 2012; Lerner et al. 2012a). Direct detection of the antigen provides earlier test results because it eliminates the delay required for the immune system to produce sufficient quantities of antibodies to be detected via Western Blot or ELISA, which can improve patient prognosis (Ma et al. 1992). Here we demonstrate that antibody-functionalized SWNT FET devices are effective biosensors for rapidly detecting Lyme flagellar antigen at clinically relevant concentrations, as low as 1 ng/ml, with negligible response to negative-control proteins and to pure buffer solution.

## 2. Materials and Methods

2.1 Device Fabrication:

Carbon nanotube transistors were fabricated using previously described methods (Khamis et al. 2011). Briefly, a solution of Iron (III) Nitrate dissolved in isopropanol was spin coated onto a p++ doped $Si/SiO_2$ wafer to give an iron catalyst layer. Single-walled nanotubes were grown by catalytic chemical vapor deposition (CVD) with methane (2500 sccm) as the carbon source in a background of forming gas (600 sccm Ar, 320 sccm $H_2$) at 900°C for 2 mins. Following the growth, an optimized bilayer photolithography process using PMGI and Shipley 1813 was used to pattern source and drain electrodes with 2.5 um channel length that were subsequently metallized (3 nm Ti/40 nm Pd) using thermal evaporation. The doped silicon substrate served as a global backgate to complete the three terminal field-effect transistor geometry. Devices were individually characterized by sweeping the back gate voltage from -10 V to 10 V while holding the bias voltage fixed at 100 mV. Approximately 80 high quality semiconducting SWNT devices with ON/OFF ratios >1000 were selected for use in experiments.

2.2 Protein functionalization

Monoclonal Lyme antibodies specific for Borrelia Burgdorferi flagellar antigen (p41) were obtained commercially (antibodies-online.com). Histidine-tagged Lyme antigen containing the p41 flagellar immunodominant region (also known as flagellin) was obtained from ProSpec. Antigen was diluted with Tris-HCl buffer (pH 7.5), aliquoted to several concentrations and stored at 4° C. Antibodies were aliquoted at a concentration of 1 µg/ml in Tris-HCl buffer and stored at -20° C.

Nanotube functionalization followed previously documented procedures (Lerner et al. 2012a) adapted from (Strano et al. 2003) (see Figure 1). Carbon nanotubes were functionalized using diazonium salts synthesized according to a published recipe (Saby et al. 1997). Samples were immersed in a solution of 4-carboxybenzene diazonium tetrafluoroborate at a concentration of 2.5 mg/1 mL deionized (DI) water for 1 hr at 40°C to create $sp^3$ hybridization sites ending in carboxylic acid groups, then rinsed for 1 min each in acetone, methanol, and deionized water baths. The carboxylic acid groups were then activated with EDC and stabilized with sulfo-NHS at concentrations of 6 mg and 16 mg per 15 mL MES buffer (pH 6.0) respectively for 15 minutes at room temperature, followed by a DI water rinse. A solution of antibodies at a concentration of 1 µg/mL was pipetted onto the nanotube devices in a humid environment to prevent the solution from evaporating, causing primary amines on the antibody to displace stabilized sulfo-NHS sites over a period of one hour. The devices were washed thoroughly by rinsing in two DI water baths for 2 mins each and dried with gentle (less than 20 psi) nitrogen flow in order to minimize the amount of salts and non-specifically bound proteins on the device.

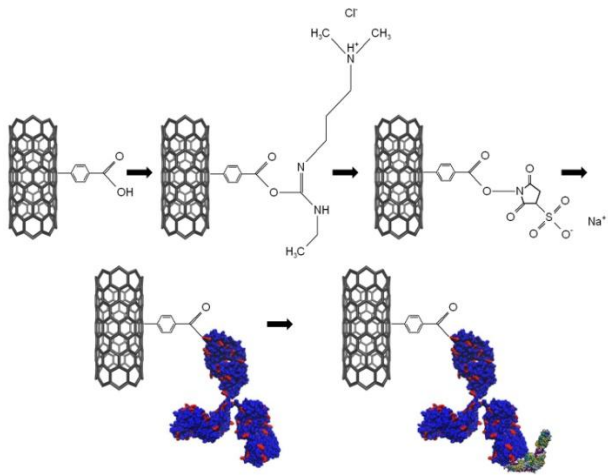

Figure 1 Functionalization chemistry for Lyme antibody and binding of flagellar antigen. First, a pristine nanotube is treated with diazonium salts to create sp³ hybridization sites ending in carboxylic acid moieties. The carboxylic acid is activated with EDC and stabilized with sulfo-NHS. The sulfo-NHS is displaced by the primary amine in a surface lysine residue (depicted in red) on the anti-p41 antibody. The flagellar antigen (depicted according to secondary structure in purple, cyan, and yellow) then binds to the epitope on the antibody during the exposure step.

Antibody-functionalized SWNT-FET devices were similarly exposed to droplets of antigen at a known concentration for 20 minutes in a humid environment to prevent the droplet from evaporating. Exposure to Lyme flagellar antigen occurred over a sufficiently long time scale to ensure adequate time for the proteins to diffuse to the sensor and establish equilibrium between bound and unbound species. This is known to be a critical consideration for detection of biomolecular analytes at pM or lower concentration (Squires et al. 2008). The devices were then washed in two DI water baths for 2 mins each to remove non-specifically bound antigen and dried with gentle nitrogen flow. Each device was exposed to only one concentration of antigen in order to avoid contamination of samples, and each concentration of antigen was tested on 5-10 functionalized devices to ensure reproducibility of the results.

## 3. Results and Discussion

In order to verify the validity of the attachment chemistry, Atomic Force Microscopy (AFM) data were gathered on an Asylum AFM in tapping mode following both antibody attachment and exposure to antigen. Figure 2a-b are AFM images that show the effectiveness of the functionalization procedure, which results in 3-5 attachment sites per micrometer of nanotube length. Statistical analysis shows that the antibodies are 2.78 nm ± 0.22 nm in size, slightly smaller than expected for a complete IgG. This is likely due to the protein being slightly distorted during tapping mode AFM in air. We conducted a control experiment to establish that nanotubes and the nanotube/$SiO_2$ interface show very low affinity for nonspecific binding of the Lyme antigen; details are provided in Supplementary Materials Figure S1. When exposed to 1 μg/mL Lyme antigen, there was minimal non-specific binding to the nanotubes or the surrounding substrate.

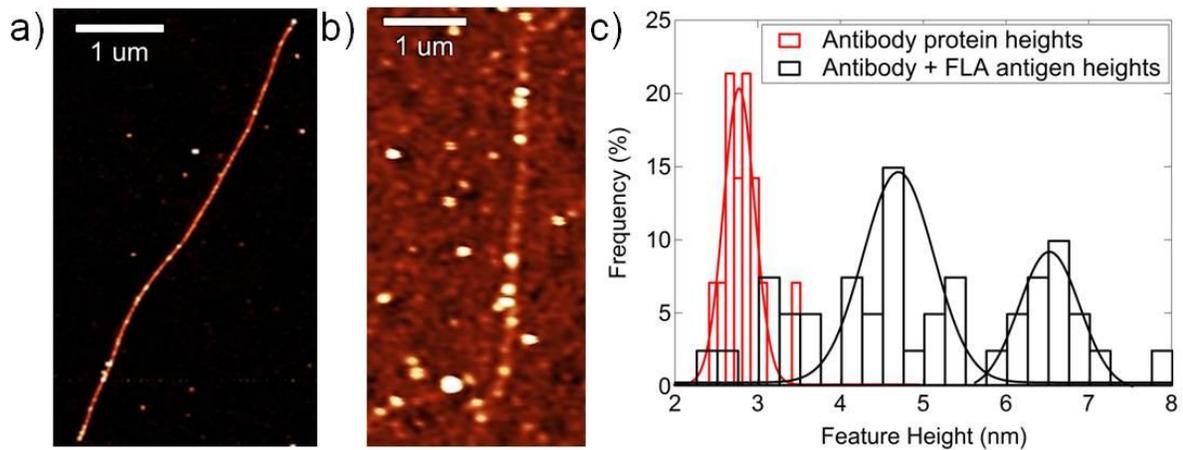

Figure 2 a) AFM image of anti-p41 antibody proteins attached to carbon nanotubes. Z scale is 6 nm, average protein feature is ~2.8 nm in size. b) AFM image after incubation in a solution of FLA antigen at a concentration of 400 ng/mL shows antigen attached to bound antibodies on a nanotube. Z scale is 8 nm . c) Histogram of feature heights following antibody attachment (red data) and exposure to antigen (black data).

After exposure to flagellar antigen at a concentration of 400 ng/mL, the histogram of feature heights shows a shift towards larger proteins complexes. There are two main peaks at feature height of 4.7 ± 0.6 nm and 6.5 ± 0.4 nm. These are associated with binding of one and two antigen proteins, respectively, to a bound antibody, consistent with the fact that the IgG used in the experiments has two binding sites (Talwar and Srivastava 2006). Each added antigen increases the feature height by approximately 1.8 nm. There also appears to be an additional peak at ~3 nm that represents unreacted antibodies; this peak accounts for ~20% of the total features measured. These data suggest that after exposure to antigen at 400 ng/mL, approximately 80% of the antibodies have bound at least one antigen, in good agreement with the electronic response data presented in Figure 3b, below.

Electronic measurements of the current as a function of the backgate voltage (I-$V_g$ characteristics) for individual NT FET devices were taken following each chemical modification to monitor the effect of chemical functionalization and to confirm attachment of antibodies (Figure 3a). Parameters of interest included $I_{ON}$, the ON state current of the device, and $V_{TH}$, the threshold voltage, where the line tangent to the I-$V_g$ curve intersects the gate voltage axis. A 50-90% drop in $I_{ON}$ as well as a 3-4 V decrease in the threshold voltage resulted from diazonium oxidation, which is associated with the creation of $sp^3$ hybridization sites terminated with a carboxylic acid group. EDC/NHS treatment resulted in a slight decrease in the ON state current and no statistically significant shift in the threshold voltage. Following antibody attachment, there was typically an increase in ON state current, suggesting that attachment led to a decrease in carrier scattering. Upon antigen exposure, a shift in the threshold voltage towards more negative values was consistently observed. There was no statistically significant change in ON state current following antigen exposure.

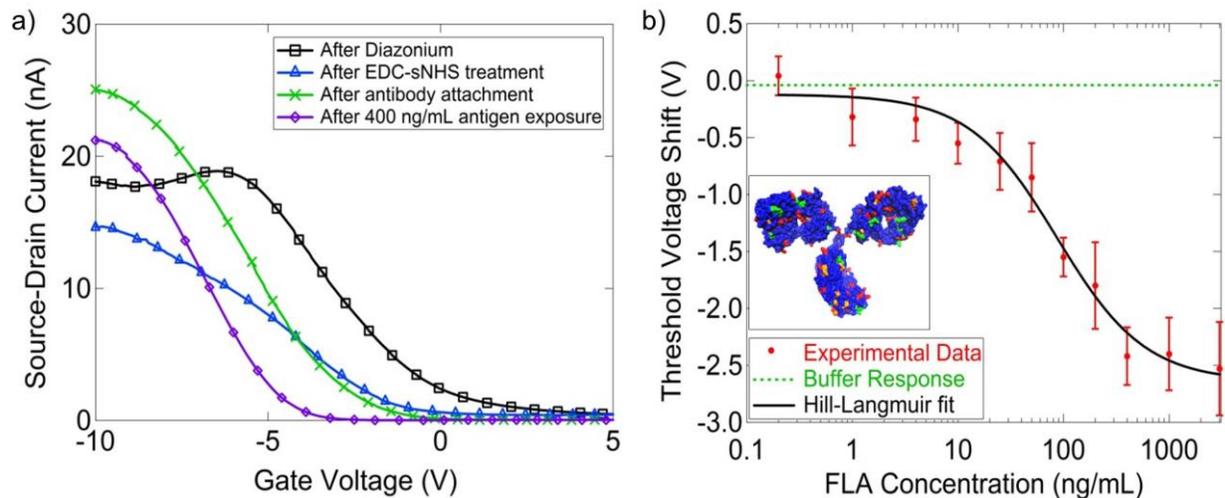

Figure 3 a) Current vs. Gate Voltage characteristics at subsequent stages of nanotube functionalization. The sensing response is the 2.3 V decrease of the threshold voltage upon exposure to antigen (green to purple curves). b) Threshold voltage shift as a function of antigen concentration can be fit with a model based on the Hill-Langmuir equation. These data indicate a limit of detection of ~ 1 ng/mL and non-cooperative antigen binding . Inset shows structure of anti-p41 Borrelia antibody with lysine (red), histidine (orange) and arginine (green) residues highlighted. Any of these basic residues could become protonated and generate a local gating effect if they come into closer contact with the nanotube upon antigen binding.

There was a systematic dependence of the threshold voltage shift with varying antigen concentration in the range 0.1 ng/ml to 3 µg/ml, with each concentration tested on 5-10 functionalized devices. The variation of the average measured shift in the threshold voltage as a function of antigen concentration is displayed in Figure 3b. Error bars shown are the standard error of the mean. The sensor responses agree with a model based on the Hill-Langmuir equation for equilibrium protein binding (see Figure 3b) (Hill 1910; Lehninger et al. 2008).

$$\Delta V_{TH} = A \frac{(c/K_d)^n}{1 + (c/K_d)^n} + Z$$

Here c is the Lyme antigen concentration, A is the sensor response at saturation when all binding sites are occupied, Z is an overall offset to account for the response to pure buffer, $K_d$ is the dissociation constant describing the concentration at which half of available binding sites are occupied, and n is the Hill coefficient describing cooperativity of binding.

The best fit to the data yielded a maximum response A = -2.52 V ± 0.32 V, offset parameter Z = -0.12 V ± 0.16 V, dissociation constant $K_d$ = 87 ng/mL ±27 ng/mL, and n = 1.04 ± 0.08. The best fit value of the offset parameter Z = -0.12 V ± 0.16 V was statistically indistinguishable from the experimentally measured responses of seven devices to pure buffer as a negative control (ΔV=0.11 V ± 0.15 V). The value of $K_d$ = 87 ± 27 ng/mL determined from the fit describes the concentration of antigen at which half the receptors are occupied; this regime coincides with antigen levels of diagnostic significance, 12-15 ng/mL (Shah et al. 2004). It is notable that the AFM image in Figure 2, taken after exposure to antigen at 400 ng/mL, showed occupation of at ~80% of the attached antibody sites, in good agreement with the prediction of the Hill-Langmuir fit that 95% of the active sites were occupied at this concentration. The slight discrepancy could be due to antibodies that were bound to the nanotube in such an orientation that their epitope is obstructed or were otherwise non-functional. The Hill-Langmuir model combined with the AFM data suggested that the number of such antibodies unavailable for binding is no more than 20% of the total bound antibodies.

The best fit value of the cooperativity parameter, n = 1.04 ± 0.08, indicates independent binding of Lyme antigen to the anti-p41 in the context of the NT-FET biosensor. The data presented in Figure 3b show that the measured responses from a collection of 5-10 devices could be used to differentiate between pure buffer solution (ΔV = 0.11 V ± 0.15 V) and buffer containing Lyme antigen at a concentration of at a concentration of 1 ng/mL (ΔV = -0.31 V ± 0.24 V). Previous work suggests that this limit of detection may be lowered by as much as a factor of 1000 by replacing the complete IgG with its single chain variable fragment, thus allowing the binding event to occur closer to the nanotube where the electrostatic influence on transport in the NT transistor will be more pronounced (Lerner et al. 2012a).

The observed reduction in turnoff voltage is consistent with a gating effect due to the presence of positive charges in the local electrostatic environment of the carbon nanotube (Heller et al. 2008). The proposed mechanism responsible for introducing these positive local charges is a conformational change in the antibody protein upon binding the flagellar antigen, which results in the carbon nanotube being exposed to different amino acids on the antibody exterior. Charges in close proximity to carbon nanotubes have been shown to shift the transistor threshold voltage by local gating (Lerner et al. 2012b). Lysine, histidine, and arginine (highlighted in the inset of Figure 3b) are common amino acids, a portion of which will be protonated when the pH is less than 7 and could thus be candidates for locally gating the nanotube as observed. Experimental conditions may be as low as pH 5 due to deprotonation of silanol groups in a thin water layer that forms on the hydrophilic $SiO_2$ surface (O'Reilly et al. 2005). Even in the absence of a quantitative understanding of the device response, the results provide strong evidence that the methods used here enabled attachment of antibody proteins to a NT-FET while maintaining both the high quality electronic characteristics of the NT device and the chemical recognition functionality characteristic of the protein.

As a further control experiment, antibody-functionalized sensors were incubated in a solution of bovine serum albumin (BSA) at high concentration (1 μg/mL) to approximate the effect of non-specific proteins present in patient samples (human serum albumin is a large component of blood plasma proteins and represents a potential interfering agent (Peters 1996)). A sample I-$V_g$ characteristic is presented in Supplementary Materials Figure S2. The sensor response to BSA, averaged over eight devices, was a shift of -0.13 V ± 0.18 V, similar to the response produced by buffer alone. We thus concluded that the Lyme flagellar antibody-functionalized carbon nanotube transistors exhibited a high level of specificity for the flagellar protein target antigen.

## 4. Conclusions

We demonstrated that antibody-functionalized SWNT FET devices are effective biosensors for rapidly detecting Lyme disease antigen at diagnostically relevant concentrations. The sensor shows a decrease in the threshold voltage in minutes rather than days, as required for a traditional clinical assay. We achieved detection of flagellar antigen protein at a concentration of 1 ng/ml with negligible response to control proteins and to pure buffer solution. The experiments showed an antigen-specific, concentration-dependent sensor response over a wide range of concentrations (1 ng/mL to 3000 ng/mL) that was in excellent quantitative agreement with a model based on the Hill-Langmuir equation of equilibrium thermodynamics. Future work includes functionalizing samples with single chain variable fragments to increase sensitivity and the use of a sensor array based on multiple antibodies in order to capture several types of proteins indicative of Lyme disease for a multiplexed biosensor platform. The rapidity and ease of use of this sensor is superior to traditional immunoassays, suggesting its utility as a point-of-care diagnostic tool.

## Acknowledgements

This research was supported by the Department of Defense US Army Medical Research and Materiel Command through grants W81XWH-09-1-0205 and W81XWH-09-1-0206 (M.L, J.D. A.T.C.J) and by the National Institutes of Health grant R01AI076342 (D.B.). We acknowledge use of facilities associated with the Nano/Bio Interface Center, National Science Foundation NSEC DMR08-32802. J.D. acknowledges support of the REU program of the Laboratory for Research on the Structure of Matter (NSF REU Site Grant DMR-1062638) and of the Penn University Scholars program.